\documentclass[%
 reprint,
 superscriptaddress,
 amsmath,amssymb,
 aps,
 prl
]{revtex4-2}

\usepackage{graphicx}
\usepackage{dcolumn}
\usepackage{bm}
\usepackage{xcolor}
\usepackage{multirow}

\begin{document}

\preprint{APS/123-QED}

\title{Dynamic Focusing to Suppress Emittance Transfer in Crab-Crossing Flat Beam Collisions}

\author{Derong Xu}\email{dxu@bnl.gov}
\author{J Scott Berg}
\author{Michael M Blaskiewicz}
\affiliation{Brookhaven National Laboratory}
\author{Yue Hao}
\affiliation{Michigan State University}
\author{Yun Luo}
\author{Christoph Montag}
\author{Sergei Nagaitsev}
\author{Boris Podobedov}
\author{Vadim Ptitsyn}
\author{Ferdinand Willeke}
\author{Binping Xiao}
\affiliation{Brookhaven National Laboratory}

\date{\today}

\begin{abstract}
Flat hadron beam collisions, though expected to enhance peak luminosity by about an order of magnitude, have not yet been demonstrated. 
Our study reveals a critical limitation: realistic fluctuations, when amplified by synchro-betatron resonance, lead to transverse emittance transfer in flat-beam collisions.
Using beam-beam simulations based on Electron-Ion Collider design parameters, we show that this effect leads to vertical emittance growth, which can distort the flat-beam profile and degrade luminosity.
We propose a dynamic focusing scheme that combines sextupoles with crab cavities to suppress the hourglass-induced resonance. This approach increases tolerance to fluctuations and improves the robustness of flat-beam collisions. This practical mitigation facilitates the adoption of flat-beam collisions in next-generation lepton-hadron colliders.

\end{abstract}

\maketitle

Electron-Ion Colliders (EICs) are powerful tools for probing nuclear structure and testing quantum chromodynamics (QCD) \cite{accardi2016electron}. After the success of HERA \cite{myers2016challenges}, which pioneered high-energy electron-proton collisions and deep inelastic scattering \cite{zeus2005measurement}, the next-generation EIC at Brookhaven National Laboratory (BNL) aims to deliver high-luminosity, polarized lepton-hadron collisions with broad energy tunability \cite{willeke2021electron}. Similar proposals, including the LHeC at CERN \cite{fernandez2012relation} and the EicC in China \cite{anderle2021electron}, reflect a global drive to push the QCD frontier. To reach high luminosity, the EIC adopts a flat hadron beam and a crab crossing scheme. These features introduce new beam dynamics challenges and motivate this study.

A large crossing angle in the interaction region (IR) is essential in colliders to avoid parasitic collisions \cite{hirata1994don}, reduce background near the interaction point (IP), and accommodate the detector and IR magnets. Although the crossing angle reduces beam overlap, the luminosity loss can be recovered by the crab crossing scheme. In this scheme, radio-frequency (RF) crab cavities apply time-dependent transverse kicks, rotating the bunches to enable effective head-on collisions \cite{palmer1988energy}. Crabbing of proton bunches has been demonstrated at CERN’s SPS \cite{calaga2021first}, and crab crossing of electron beams at KEKB enabled record luminosities in lepton colliders \cite{abe2007compensationof}. As a result, crab crossing has become a key feature in modern collider designs, including HL-LHC \cite{apollinari2017high}, EIC, and EicC. Besides restoring geometric luminosity, crab cavities also suppress low-order synchro-betatron resonances from the crossing angle. The EIC baseline design includes both first- and second-order harmonic crab cavities in the hadron ring \cite{xu:ipac2021-wepab009}.

Luminosity is the primary figure of merit for a collider, and beam-beam interactions often limit its achievable value, as observed in both lepton \cite{PhysRevLett.92.214801} and hadron machines \cite{PhysRevSTAB.18.121003}. The beam-beam parameter, which characterizes the maximum tune shift induced by collisions, quantifies the strength of the electromagnetic interaction between the opposing beams. 
Defining beam flatness as the aspect ratio at the IP, $\kappa\equiv \sigma_y^*/\sigma_x^*$,
the luminosity and beam-beam parameters for Gaussian-distributed colliding beams can be expressed as \cite{chao2023handbook}:
\begin{equation}
    L\propto \frac{1}{\kappa},\quad
    \xi_x\propto \frac{\beta_x^*}{1+\kappa},\quad
    \xi_y\propto \frac{\beta_y^*}{\kappa(1+\kappa)}
\end{equation}
where $\beta^*_{x,y}$ denote the $\beta$ functions at the IP, $L$ is luminosity, and $\xi_{x,y}$ are 
beam-beam parameters. 
By scaling $\beta_x^*\propto 1+\kappa$ and $\beta_y^*\propto \kappa(1+\kappa)$,
one can increase luminosity with a smaller $\kappa$ while keeping the beam-beam parameters constant.

The successful demonstration of flat hadron beam at RHIC \cite{PhysRevLett.132.205001} supports the adoption of a flat beam configuration in the EIC baseline, which aims to achieve a peak 
luminosity of $10^{34}~\mathrm{cm}^{-2}\mathrm{s}^{-1}$.
The EIC design chooses the flatness
$\kappa=\beta_y^*/\beta_x^*=\epsilon_y/\epsilon_x<0.1$, where $\epsilon_{x,y}$ are the transverse beam emittances.
The resulting small vertical $\beta_y^*$, in both hadron and electron rings, becomes comparable to the hadron bunch length $\sigma_z$, 
giving rise to the hourglass effect \cite{furman1991hourglass}.
Although this effect does not significantly reduce luminosity, it introduces nonlinear 
longitudinal-transverse coupling that excites higher-order synchro-betatron resonances 
\cite{PhysRevAccelBeams.24.041002}. 

\begin{figure}
    \centering
    \includegraphics[width=\columnwidth]{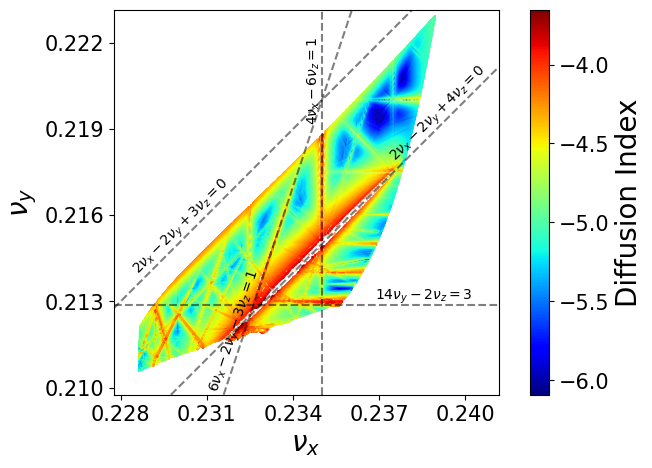}
    \caption{Frequency map showing the resonance $2\nu_x-2\nu_y+p\nu_z=0$, induced by the hourglass effect. Particles span transverse planes $(0,6\sigma_x^*)\times (0, 6\sigma_y^*)$,
    with a longitudinal offset $z=3\sigma_z$.}
    \label{fig:fmaReference}
\end{figure}

As shown in Fig. \ref{fig:fmaReference}, frequency map analysis reveals 
a clear resonance line at 
\begin{equation}
2\nu_x-2\nu_y+p\nu_z=0
\end{equation}
where $p$ is an integer and $\nu_{x,y,z}$ are
horizontal, vertical and longitudinal tunes, respectively.

When beam dynamics is governed by this resonance, there are apparently two invariants \cite{PhysRevE.49.5706}:
\(J_x + J_y = \text{const}\) and \(2J_z + pJ_y = \text{const}\), where \(J_{x,y,z}\) are the action variables
in all three planes. In synchrotrons, the longitudinal emittance typically 
exceeds the transverse emittances by several orders of magnitude. As a result, most particles satisfy 
\(J_z \gg J_{x,y}\). For these particles, when they are off resonance, they exhibit 
small-amplitude oscillations 
in \(J_y\). When on resonance, the vertical action oscillates around the center \((J_x + J_y)/2\), leading 
to transverse emittance exchange. Since the beam-beam force weakens at large amplitudes, the tune shift 
always dominates the resonance width \cite{chao1986nonlinear}. The hourglass effect further modulates the transverse tunes with respect to
longitudinal position, so particles drift in and out of resonance. 
As a result, particles eventually reach a modified transverse equilibrium distribution.

\begin{figure}
    \centering
    \includegraphics[width=\columnwidth]{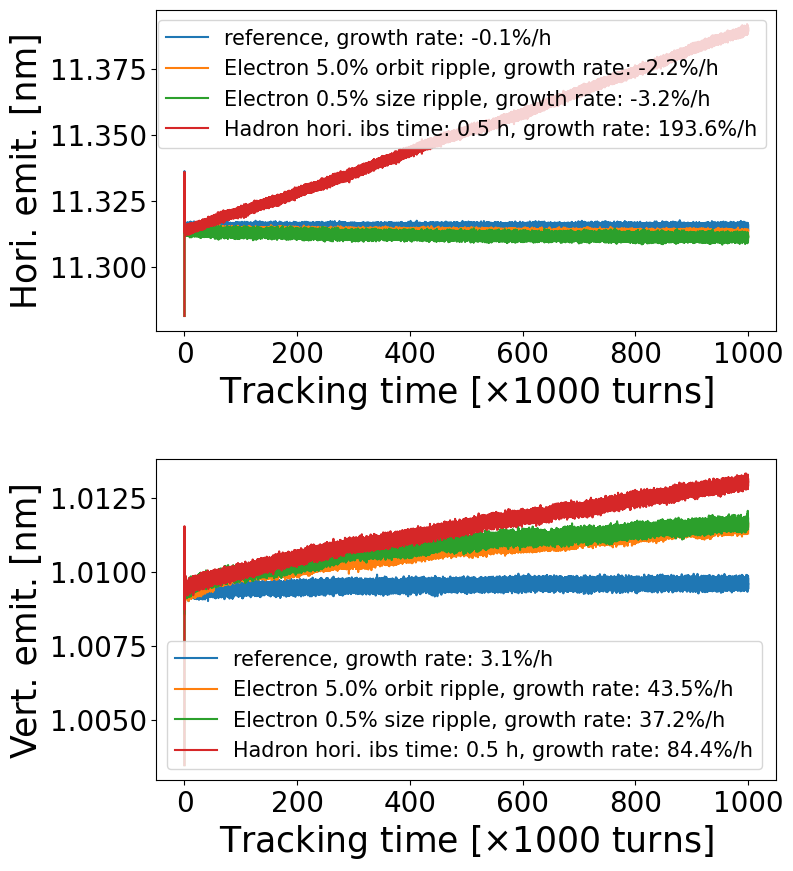}
    \caption{Emittance evolution under three types of fluctuations: electron orbit ripple, electron beam size ripple, and hadron intrabeam scattering (IBS). Orbit and size ripples are modeled as narrow-band signals centered at $60~\mathrm{Hz}$ with a $10~\mathrm{Hz}$ bandwidth, representative of power supply noise. Their RMS amplitudes are normalized to the electron beam size at the IP. IBS is modeled as a wide-band stochastic process, with strength defined by the $1/e$ horizontal emittance growth time.}
    \label{fig:emitGrowth}
\end{figure}

However, in real-world accelerator operations, physical noise is unavoidable. External diffusion can be significantly enhanced near a resonance, a phenomenon known as resonance streaming \cite{tennyson1982resonance}. The small emittance ratio $\epsilon_y/\epsilon_x$ further increases the susceptibility of the vertical emittance to such fluctuations.

The physical noise sources vary widely in amplitude and frequency spectrum. For simplicity, this letter examines robustness against three representative types: low-frequency electron orbit ripple, electron size ripple, and hadron intrabeam scattering (IBS). The orbit and size ripples are primarily driven by magnet power supply noise and typically exhibit narrow-band spectra well separated from the hadron betatron tunes \cite{podobedov2023eddy}. In contrast, IBS acts as a broadband, white-noise source that overlaps with the hadron tune lines, though its diffusion amplitude is much smaller.

Figure~\ref{fig:emitGrowth} shows significant vertical emittance growth under all three types of fluctuations. For the electron orbit and size ripples, the horizontal emittance decreases over time, indicating transverse emittance transfer. In the IBS case, the horizontal emittance growth rate is reduced from the nominal $200\%/\mathrm{h}$ to a fitted $193\%/\mathrm{h}$, again reflecting redistribution to the vertical plane. Compared with the orbit ripple, the electron size ripple causes greater vertical growth at the same magnitude, as it induces both direct diffusion and tune modulation, enhancing resonance crossing and amplifying emittance exchange.

Since practical cooling for high-energy hadron beam is still under development \cite{annurev:/content/journals/10.1146/annurev-nucl-102313-025427, PhysRevLett.102.114801,Jarvis2022}, vertical emittance growth from resonance streaming presents a distinct challenge. Emittance tends to transfer from the plane with larger emittance to the one with smaller emittance. Unlike dispersion-based mechanisms that redistribute diffusion or cooling rates \cite{wei2001intra, PhysRevAccelBeams.22.081003}, synchro-betatron resonance induces a unidirectional emittance flow without enabling cross-plane cooling. This necessitates dedicated vertical cooling: vertical growth driven by horizontal IBS cannot be suppressed by horizontal cooling alone.

Unequal transverse emittances, synchro-betatron resonance, and physical fluctuations together form the essential ingredients for driving emittance transfer. One mitigation strategy is to reduce the fluctuation amplitudes from known sources, such as magnet power supply ripple \cite{podobedov:ipac2025-mops065}. {While effective in principle, this approach increases the cost of accelerator subsystems or even require technologies beyond the current state of the art. Dedicated feedback systems could alleviate such constraints, but unknown or unpredictable noise sources during machine operation pose an ongoing challenge}. These considerations highlight the need for an active strategy to suppress emittance transfer.

Another possible mitigation strategy is to avoid resonance lines within the beam-beam tune footprint. However, for the EIC, the pursuit of high luminosity necessitates a large beam-beam footprint. In addition, the short hadron bunch length and constraints on the RF system together limit the available range of the longitudinal tune. As a result, the resonance condition discussed earlier becomes unavoidable. Since this resonance originates from the hourglass effect, it can be weakened by leveraging the concept of dynamic focusing.

Originally proposed by R. Brinkmann for THERA \cite{brinkmann1995method}, the dynamic focusing scheme employs RF quadrupoles (RFQs) to generate time-dependent focusing. Particles receive varying focusing based on their longitudinal positions, aligning their beta functions at the collision point (CP). While the concept has been considered to mitigate head-tail-like instabilities from strong hourglass effects \cite{WangBB24hourglass} and mentioned as a potential option for future colliders \cite{acar2017future}, {designing RFQs to provide such dynamic focusing remains technically challenging. No implementation has yet been demonstrated in an operational machine}.

The crab-waist scheme offers an effective implementation of dynamic focusing. Crab sextupoles apply a specific transformation to suppress beam-beam resonance \cite{PhysRevSTAB.14.014001}. A side effect of this transformation is the redistribution of the vertical beta function within the overlap region, yielding a modest geometric luminosity gain \cite{PhysRevLett.104.174801}.

A comparable dynamic focusing effect to cancel the betatron modulation can be achieved by combining sextupoles and crab cavities. The CP location in the crab crossing scheme remains determined by the longitudinal coordinates of the colliding particles in opposite beams. The hourglass effect, manifested as vertical betatron modulation, arises from the drift space between the IP and CP.  Since the hadron bunch is much longer than the electron bunch, the CP is effectively located at a distance $S\approx z/2$ from the IP \cite{hirata1992symplectic, PhysRevAccelBeams.27.061002}, where $z$ is the longitudinal coordinate of the hadron particle. The crab cavity imparts a longitudinally varying transverse kick, characterized by the crab dispersion \cite{PhysRevAccelBeams.25.071002}. Downstream of the cavity, this kick translates into a horizontal displacement. A sextupole placed in this region can provide a time-dependent focusing, effectively counteracting the IP-to-CP shift. This mechanism resembles the chromaticity compensation scheme in the final focusing section \cite{PhysRevLett.86.3779}, though it relies on nonzero crab dispersion rather than momentum dispersion.

\begin{figure}
    \centering
    \includegraphics[width=\columnwidth]{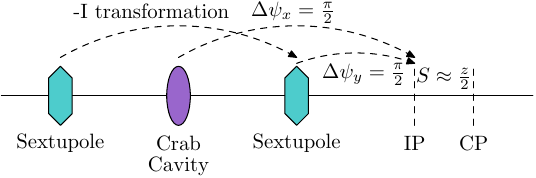}
    \caption{Locations of sextupoles and crab cavities to generate dynamic focusing. The other side of the IP is omitted for clarity, assuming symmetry and a locally closed crab crossing scheme. The values of $\Delta \psi_{x,y}$ represent the remainders of the corresponding phase advances modulo \(\pi\).}
    \label{fig:sext}
\end{figure}

Figure~\ref{fig:sext} illustrates the relative position in a crab crossing scheme to generate dynamic focusing. A horizontal local crab crossing scheme is assumed, in which the crab dispersion is confined between the upstream and downstream crab cavities around the IP. The horizontal phase advance 
between crab cavities and the IP is $\pi/2~\mathrm{mod}~\pi$.

To compensate for the vertical IP-to-CP shift, a sextupole is inserted between the crab cavity and the IP
with the vertical phase advance of $\pi/2~\mathrm{mod}~\pi$. The required integrated sextupole strength is given by:
\begin{equation}
K_2L=\frac{\sqrt{{\beta_x^*}/{\beta_{s,x}}}}{4\theta_c\beta_{s,y}\beta_y^*\cos\psi_x}
    \label{eq:sextupoleStrength}
\end{equation}
Here, $\theta_c$ is the half crossing angle, $\psi_x$ is the horizontal phase advance from the sextupole to the IP, and $\beta_{s,x}$ and $\beta_{s,y}$ are the beta functions at the sextupole.

To cancel the second-order geometric terms introduced by the sextupole, another sextupole is placed outside the crab cavities, such that the two sextupoles form a $-I$ transformation. This cancellation is imperfect because the crab dispersion vanishes at the second sextupole. As a result, the residual horizontal kick at the CP is:
\begin{equation}
    \Delta p_x=-\frac{\beta_{s,x}}{\beta_x^*\beta_{s,y}\beta_y^*}\left(\frac{\theta_c z_0^2}{4}+\frac{x_0z_0}{2}\right)
    \label{eq:residual}
\end{equation}
where $x_0$ and $z_0$ are the horizontal and longitudinal phase space coordinates at the IP. This residual term can be minimized by reducing the ratio $\beta_{s,x}/\beta_{s,y}$, thereby ensuring that the perturbation to horizontal dynamics is negligible.

\begin{figure}
    \centering
    \includegraphics[width=\columnwidth]{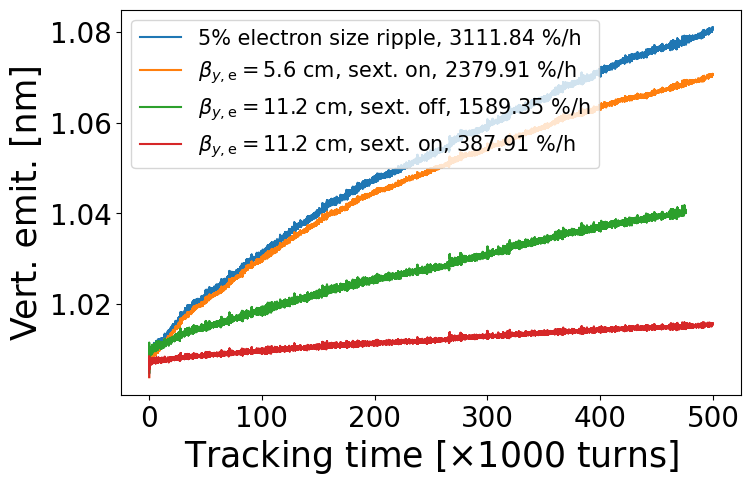}
    \caption{Hadron vertical emittance evolution under a $5\%$ electron beam size fluctuation, with and without dynamic focusing and increased electron vertical beta function.}
    \label{fig:sextTest}
\end{figure}

Figure~\ref{fig:sextTest} demonstrates the effectiveness of the sextupole correction scheme under a $5\%$ electron beam size fluctuation, which otherwise induces substantial vertical emittance growth. In addition to the hadron beam’s vertical beta function, the electron vertical beta function $\beta_{y,\mathrm{e}}^*$ also contributes to the hourglass effect, as it governs the variation of the electron beam size during the beam-beam interaction. While doubling $\beta_{y,\mathrm{e}}^*$ or applying sextupole correction individually provides partial mitigation, their combination yields a pronounced suppression of emittance growth. These results reinforce that the vertical emittance growth originates from the hourglass effect and confirm the efficacy of the proposed correction strategy.

The $\beta_{y,\mathrm{e}}^*$, however, is also constrained by beam-beam interactions. In a linac-ring collider, the condition $\beta_{y,\mathrm{e}}^*>\sigma_z$ is more readily satisfied. In contrast, in a ring-ring collider, $\beta_{y,\mathrm{e}}^*$ is limited by the maximum sustainable beam-beam parameter; exceeding this threshold leads to electron vertical emittance blow-up. Simulations indicate an upper limit around $0.15$ \cite{PhysRevSTAB.7.104401}, while typical values achieved in circular lepton colliders operations are slightly above $0.1$ \cite{Shatunov2016,Zhou_2024}. Beam-beam studies for the EIC are consistent with these limits. Strong-strong simulations show that the electron beam emittance remains stable for $5.6~\mathrm{cm}<\beta_{y,\mathrm{e}}^*<8.4~\mathrm{cm}$, corresponding to a vertical beam-beam parameter of $0.10–0.15$. Our study favors operating at a larger $\beta_{y,\mathrm{e}}^*$, within this feasible range, to suppress the impact of hourglass effect on the hadron beam.

Implementing the dynamic focusing scheme with an increased $\beta_{y,\mathrm{e}}^* = 7.2~\mathrm{cm}$ improves tolerance to electron orbit ripple, size fluctuation, and hadron IBS diffusion by factors of two to three, accompanied by a slight gain in luminosity. This enhanced robustness against real-world noise underscores the scheme’s practical advantage.

\begin{figure}[h]
    \centering
    \includegraphics[width=\columnwidth]{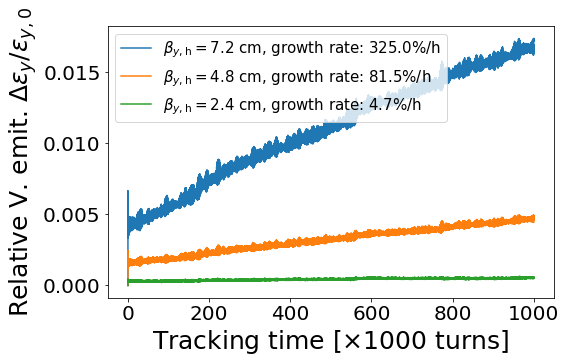}
    \caption{Hadron vertical emittance evolution under a $5\%$ electron beam size ripple with dynamic focusing. Emittance is normalized to its initial value.
    Reducing hadron $\beta_y^*$ to $2.4~\mathrm{cm}$ eliminates vertical emittance growth. In contrast, the same fluctuation causes a growth rate exceeding $3000\%/\mathrm{h}$ without dynamic focusing (see Fig.~\ref{fig:sextTest}).}
    \label{fig:reducedBeta}
\end{figure}

Beyond the EIC, this method offers broader applicability to future lepton–hadron colliders. 
The EIC design constrains vertical divergence to match the horizontal at the IP \cite{ptitsyn:ipac2023-wezg1}, limiting hadron vertical
beta function $\beta_{y,\mathrm{h}}^*$ reduction. Dynamic focusing decouples $\beta_{y,\mathrm{h}}^*$ from
the hadron bunch length $\sigma_z$. By reducing $\beta_{y,\mathrm{h}}^*$ while holding $\sigma_y^*$ constant, the emittance ratio $\epsilon_y/\epsilon_x$ improves, the vertical beam-beam parameter decreases, and resonance driving term strengths diminish. Simulation results in Fig.~\ref{fig:reducedBeta} show that, at a vertical beta of $2.4~\mathrm{cm}$, even a $5\%$ electron beam size fluctuation induces no vertical emittance growth. This strategy opens new opportunities for robust flat hadron beam collisions in future high-luminosity colliders.

However, achieving the required beam optics poses notable challenges. The crab cavities demand a specific horizontal phase advance and a large horizontal beta function to compensate for the crossing angle, whereas the sextupole requires precise vertical phase control relative to the IP. Sextupole placement further benefits from a large vertical beta function and a small horizontal beta function to reduce both its required strength and residual nonlinear effects. In the EIC-like local crabbing scheme, the proximity of crab cavities to the final focusing quadrupoles limits available space for additional sextupoles due to layout constraints. 
Nonetheless, these design challenges appear manageable within the scope of collider optics.

One possible solution is to place the crab cavities further away from the IP, allowing better separation of the horizontal and vertical beta peaks and providing more degrees of freedom to control phase advances. An alternative is the global crabbing scheme, successfully operated at KEKB and considered for the HL-LHC \cite{PhysRevSTAB.13.031001}. It has also been evaluated for the EIC, with no adverse beam-beam effects identified \cite{xu:ipac2024-mopc71}. Its potential to facilitate dynamic focusing merits further exploration. 

In summary, we identify emittance transfer in flat-beam collisions, driven by synchro-betatron resonance that amplifies physical fluctuations. To suppress this effect, we propose a dynamic focusing scheme combining sextupoles with crab cavities. This approach improves noise tolerance, enables a modest luminosity gain, and offers a practical path to relax the hourglass constraint in future colliders.

\begin{acknowledgments}
This research used resources of the National Energy Research Scientific Computing Center (NERSC), a U.S. Department of Energy Office of Science User Facility operated under Contract No. DE-AC02-05CH11231.
This work was supported by Brookhaven Science Associates, LLC under Contract No. DE-SC0012704 with the U.S. Department of Energy, and by a U.S. Department of Energy Early Career Award.

\end{acknowledgments}



\bibliography{ref}

\end{document}